# ClausewitzGPT Framework: A New Frontier in Theoretical Large Language Model Enhanced Information Operations

By Ben Kereopa-Yorke, MCybSec
UNSW Canberra at the Australian Defence Force Academy
October, 2023

**Abstract**

In a digital epoch where cyberspace is the emerging nexus of geopolitical contention, the melding of information operations and Large Language Models (LLMs) heralds a paradigm shift, replete with immense opportunities and intricate challenges. As tools like the Mistral 7B LLM (Mistral, 2023) democratise access to LLM capabilities (Jin et al., 2023), a vast spectrum of actors, from sovereign nations to rogue entities (Howard et al., 2023), find themselves equipped with potent narrative-shaping instruments (Goldstein et al., 2023). This paper puts forth a framework for navigating this brave new world in the "ClausewitzGPT" equation. This novel formulation not only seeks to quantify the risks inherent in machine-speed LLM-augmented operations but also underscores the vital role of autonomous AI agents (Wang, Xie, et al., 2023). These agents, embodying ethical considerations (Hendrycks et al., 2021), emerge as indispensable components (Wang, Ma, et al., 2023), ensuring that as we race forward, we do not lose sight of moral compasses and societal imperatives.

$$ClausewitzGPT(P, S, T)$$
$$= h\left(f_{LLM}(P), \sum_{i=1}^{n} f_{agent_i}(f_{LLM}(P), S, T), S, T\right) \times \omega_{machine-speed}$$

Mathematically underpinned and inspired by the timeless tenets of Clausewitz's military strategy (Clausewitz, 1832), this thesis delves into the intricate dynamics of AI-augmented information operations. With references to recent findings and research (Department of State, 2023), it highlights the staggering year-on-year growth of AI information campaigns (Evgeny Pashentsev, 2023), stressing the urgency of our current juncture. The synthesis of Enlightenment thinking, and Clausewitz's principles provides a foundational lens, emphasising the imperative of clear strategic vision, ethical considerations, and holistic understanding in the face of rapid technological advancement.

Given this backdrop, the need for interdisciplinary insights from computational ethics, computational social science, and systems engineering becomes paramount (Evgeny Pashentsev, 2023). These disciplines offer the tools and perspectives necessary to navigate the complexities of the digital age (Bhaso Ndzendze & Tshilidzi Marwala, 2023), ensuring that AI-augmented operations not only achieve strategic objectives but do so with a profound respect for ethical boundaries and societal implications (Richard Ned Lebow, 2007).

**Introduction**

Throughout history, the strategic dissemination and manipulation of information have played pivotal roles in shaping societies, politics, and conflicts (Goldstein et al., 2023). From the poignant speeches of ancient Greek orators to the strategic artistry of Roman murals, the art of information operations has thrived and evolved (Goldstein et al., 2023). Over the centuries, technological breakthroughs like the printing press, radio, and television continually reinvented this art. Yet, the 21st-century digital metamorphosis, characterised by the internet's ubiquity and the pervasive influence of social media, catapulted these dynamics into an unprecedented realm (Pashentsev, 2023).

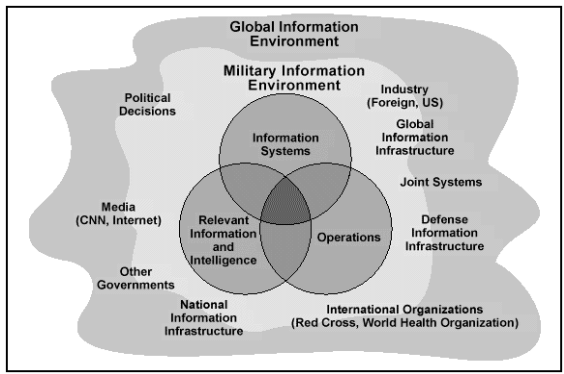

*Figure 1: Information Operations Components, FM-100-6, Department of the Army, 1996*

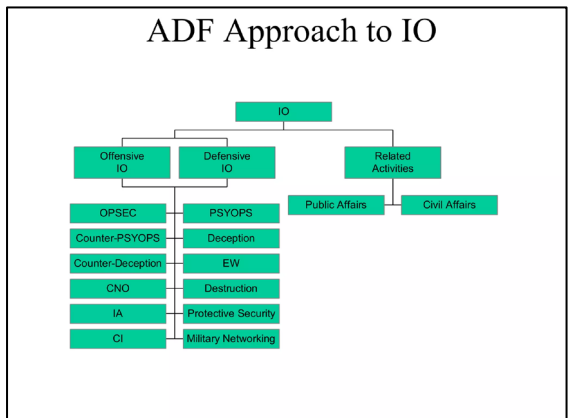

*Figure 2: Information Operations in the Australian Defence Force, Captain J. Malone, 2004*

In this ever-evolving panorama of information operations, Artificial Intelligence (AI) emerges as the next frontier. AI is a branch of computer science that seeks to create machines capable of mimicking human cognitive functions. It's not about merely executing tasks that a computer can already do, but rather, it's about enabling machines to perform tasks that, until now, only humans could do. This includes abilities like understanding natural language, recognizing patterns, making decisions based on data, and learning from new information. Over the past few decades, AI has moved from being a speculative science fiction trope to a tangible, impactful reality, transforming industries and societal structures along the way.

With its inherent capability to process vast amounts of data, discern patterns, and adapt in real-time, AI offers a toolkit that magnifies the efficiency and reach of information operation campaigns (Goldstein et al., 2023). Modern nations and entities, recognising the value of AI, are incrementally integrating it into their strategies (Department of State, 2023), harnessing its power for both benevolent and malevolent purposes. Algorithms, while aiming to optimise user experiences, inadvertently risk creating myopic digital echo chambers, thereby amplifying specific narratives and potentially sidelining others.

Recognising AI's potential, modern nation-states are progressively assimilating it into their strategic repertoires, wielding its power for both benevolent and nefarious ends (Pashentsev, 2023).

Drawing inspiration from recent research (Goldstein et al., 2023), this thesis endeavours to chart the conceivable trajectory of LLMs in information operations. Viewed from a futurist lens, the integration of LLMs into information operations seems not just likely but inevitable. As AI-driven tools become staples in information warfare (Murphy, 2023), LLMs could emerge as the vanguard, bringing unprecedented scalability, adaptability, and precision. Envision a scenario where information campaigns are tailored dynamically to individual psychological profiles, where messages evolve in real-time based on audience feedback, and where narratives are crafted with a nuance and depth that were hitherto the exclusive domain of human intellect. Such a future, while offering immense opportunities, also presents profound challenges.

The ethical ramifications of LLM-driven operations, the risk of deepening misinformation chasms, and the overarching need for transparency and accountability are issues that will demand attention (Raska & Bitzinger, 2023). The conceptual "ClausewitzGPT" model, which will be elaborated upon in this thesis, offers a pioneering framework to navigate this uncharted territory. Merging traditional strategic thinking with the capabilities of LLMs, it seeks to balance the promise of technological innovation with the imperatives of ethical responsibility.

**Nation-State Advances in AI-driven Information Operations**

In the realm of information operations, recent advancements have highlighted the increasingly complex landscape, intertwined with technology, politics, and national interests. Nation-states are continually evolving their strategies, leveraging modern tools like AI to achieve their objectives (Scharre, 2023).

A salient example is the series of datasets recently released by Twitter (Twitter, 2021)., which unveiled an extensive network of accounts attributed to state-backed operations within the People's Republic of China (PRC). These accounts, operating through VPNs and even some direct IP addresses from mainland China, were primarily designed to influence the Chinese diaspora and international communities (Wallis et al., 2022). The narratives they disseminated spanned a broad spectrum - from domestic policies to international events, from deriding Chinese dissidents to scrutinising US management of the Covid-19 pandemic. The utilisation of multimedia content, from images to videos, was a hallmark of this campaign. Provocative imagery and dense textual content sought to shape perspectives on matters like the Hong Kong protests and the CCP's handling of the pandemic. Yet, a closer look reveals a rather hasty assembly of content, sometimes even betraying its machine-generated nature with visible editing tools. Furthermore, the attempt to impersonate Western users, evident through the adoption of common Western names and mismatched activity timings, exemplified the campaign's broader intent but also its shortcomings (Twitter, 2021).

While the efficacy of these operations remains a subject of debate (Howard et al., 2023), there's an undeniable indication of nation-states experimenting with methodologies. For instance, by February 2021, traces of similar operations had surfaced in media outlets across Greece and India, hinting at an evolving sophistication in tactics (Pashentsev, 2023).

Further affirming the growing role of the PRC in the information warfare landscape, the US Department of State unveiled a comprehensive report (Department of State, 2023), detailing the PRC's endeavours to mold the global information environment. This report highlighted Beijing's massive investments in establishing a global ecosystem favourable to its propaganda, while also promoting censorship and disseminating disinformation. The multifaceted strategy encompassed propaganda, leveraging digital authoritarianism, exploiting international partnerships, and even exerting influence over Chinese-language media.

The repercussions of these efforts are vast and varied. From influencing content on popular news sites to reshaping the global digital television landscape, the PRC's footprint is evident. Its ventures into promoting digital authoritarianism further underscore the strategic deployment of technology for geopolitical gains (Mehdi Parvizi Amineh, 2022). The spread of PRC's digital ecosystems, such as "smart" cities and other surveillance tools, across various countries, is just a glimpse of this growing influence (Mehdi Parvizi Amineh, 2022).

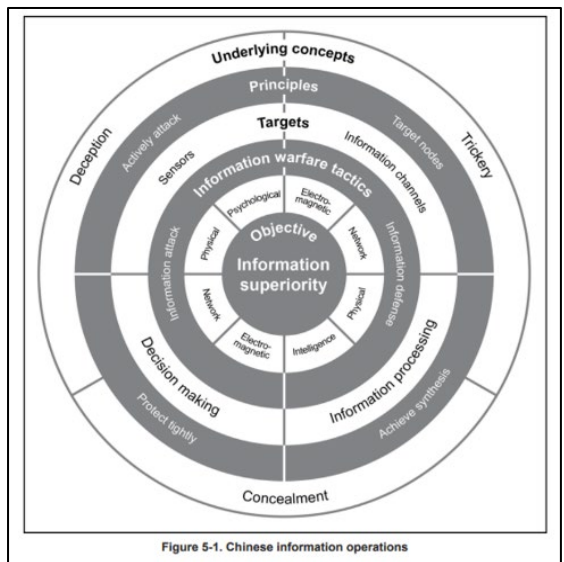

*Figure 3: Chinese Information Operations, APAN, 2021*

However, the "The Palgrave Handbook of Malicious Use of AI and Psychological Security" (Pashentsev, 2023), offers a broader perspective, emphasising the ubiquitous threat posed by AI-enabled disinformation campaigns. By 2020, a staggering 81 countries had witnessed the use of social media platforms for spreading computational propaganda. This surge, from just 28 countries in 2017, underscores the escalating challenge. The potential for AI to manipulate public opinion by analysing sentiments in social media posts further accentuates the need for robust defences and governance.

### Theoretical Impact of LLMs on Information Operations

Within the vast realm of AI, Large Language Models, or LLMs, represent a particularly intriguing and advanced subfield. These models are trained on enormous datasets containing vast amounts of text from diverse sources (Vaswani et al., 2017). This training enables them to not just understand language but to generate it in ways that are indistinguishably human-like. An LLM doesn't just respond to inputs; it can craft narratives, generate creative content, answer questions, and even engage in complex debates. The sophistication of these models, exemplified by models like GPT-4, has led many to envision them as the next frontier in AI, where the line between machine-generated and human-generated content becomes increasingly blurred.

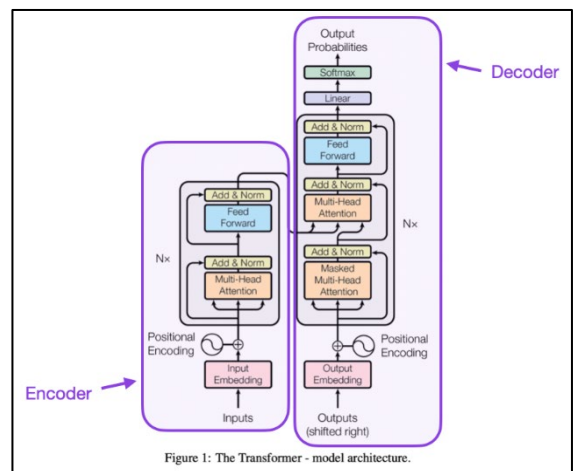

*Figure 4: Transformer Model Architecture, Vaswani et al., 2017*

At the heart of the LLM's power lies the Transformer architecture (Vaswani et al., 2017), as depicted above. This architecture revolutionises the way machines understand and generate language. Comprising multi-headed attention mechanisms, the Transformer meticulously weighs the importance of different words in a sentence, capturing context with remarkable accuracy. The stacked layers of encoders and decoders enable it to process input text and produce coherent, contextually relevant outputs. Such deep and intricate processing allows the LLMs to craft narratives, answer queries, and even engage in simulated human-like debates.

The capabilities of LLMs extend far beyond chatbots or automated customer service tools (Jin et al., 2023). When applied to the domain of information operations, their potential becomes both fascinating and formidable.

Here are some prospective applications:

- **Real-time Adaptation**: Based on audience feedback and engagement metrics, LLMs could adapt messaging strategies in real-time, refining content to achieve desired outcomes.
- **Simulation and Forecasting**: LLMs could simulate potential responses to various narratives, helping strategists forecast public reactions and adjust campaigns accordingly.
- **Multilingual Operations**: With their vast linguistic capabilities, LLMs could operate across multiple languages, ensuring consistent messaging across diverse audiences.
- **Deep Analysis**: By sifting through vast amounts of data, LLMs can identify trends, sentiments, and emerging narratives, providing invaluable insights to information operation strategists.

However, it's crucial to approach these applications with a sense of caution and responsibility. The power of LLMs in shaping perceptions and influencing public opinion is immense. Ensuring they are used ethically and transparently is of paramount importance as we step into this new era of information operations.

In the intricate arena of information operations, understanding foundational tenets is paramount (Haig, 2020). At its core, information operations encompass activities strategically designed to influence, disrupt, or guide decision-making processes, often transcending borders and reaching global audiences (Department of the Army, 1996). Different nations and actors have varying approaches to these operations, influenced by their unique geopolitical, technological, and sociocultural contexts.

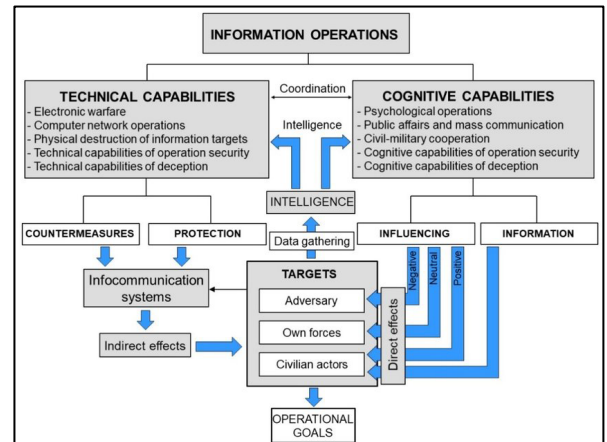

*Figure 5: Information capabilities of information operations and their effects, Haig, 2020*

Traditional methods, which often grapple with the lag between message creation and feedback integration, find a potential solution in LLMs. Equipped with cutting-edge analytics, these models can recalibrate their strategies in real-time, based on audience reactions, thus ensuring that campaigns consistently hit their mark. The narratives they weave, while potentially compelling, also usher in ethical dilemmas. Questions arise about the boundaries of crafting persuasive narratives, especially when the line between accuracy and manipulation becomes nebulous. Who arbitrates the balance between strategic objectives and ethical considerations?

Furthermore, while LLMs excel at emulating human language, they inherently lack the nuanced tapestry of genuine human emotions and cultural sensitivities (Goldstein et al., 2023). This could lead to content that, albeit technically sound, might resonate as dispassionate or even discordant. As nations and entities recognise the potential of LLMs, there's also the looming risk of over-dependence, potentially sidelining human-centric strategic innovation and introducing vulnerabilities to adversarial interventions (Goldstein et al., 2023).

**"ClausewitzGPT" and Modern Strategy**

Carl von Clausewitz's profound insights into the nature of conflict and the interplay of war and politics have shaped generations of strategic thought (Wallace, 2018). In "On War," (Clausewitz, 1832) he delved into the intricacies of war, emphasising its chameleon-like nature. As we stand on the cusp of a new era of information warfare, it is only just that we turn to Clausewitz for guidance, adapting his principles to our modern challenges (Wallace, 2018). Enter the "ClausewitzGPT" model.

This model, a fusion of traditional warfare strategy and state-of-the-art AI technologies, endeavours to craft a framework for AI-augmented information operations that is both potent and principled. Drawing from the Generative Pre-trained Transformers architecture that underlies modern LLMs (Vaswani et al., 2017), "ClausewitzGPT" is more than just a tool—it represents a strategic doctrine for the digital age.

Clausewitz emphasised that the most far-reaching act of judgment that the statesman and commander must make is to establish the kind of war on which they are embarking (Clausewitz, 1832). Similarly, when deploying LLMs in information operations, the foremost act is to define the nature and objectives of the campaign. The LLM, in this context, is not just a tool but part of a process chain, integrating seamlessly into the larger strategic ecosystem, ensuring alignment with overarching objectives.

However, echoing Clausewitz's sentiments on the 'fog of war' (Clausewitz, 1832), the realm of digital information operations is fraught with uncertainties. The dynamism of algorithms, the unpredictability of their outputs, and the vast, ever-shifting digital battleground can create a 'fog' that obscures clear decision-making. But just as Clausewitz advocated for a deep understanding of the enemy and the environment (Clausewitz, 1832), so too must modern strategists understand the intricacies of LLMs and the digital terrain.

The "ClausewitzGPT" model also serves as a cautionary framework. The seductive power of LLMs can lead to an overemphasis on the medium rather than the message. Clausewitz's assertion that war should always serve political ends, and not vice versa (Clausewitz, 1832), is a vital reminder in this context. LLM-driven operations, while powerful, must always be subservient to the broader strategic and political objectives.

**Mathematical Foundations**

These equations aspire to numerically delineate the efficacy and dynamics of information operations, bridging the gap from traditional strategies to the contemporary realm of AI-enhanced techniques.

Starting with the representation of traditional information operations (Department of the Army, 1996), the below equation is proposed:

$$\text{Impact}_{\text{Traditional}} = \frac{\text{Reach} \times \text{Message Quality} \times \text{Engagement}_{\text{Traditional}}}{(\text{Media Noise} + \text{Resistance} + \text{External}_{\text{factors}}) \times \text{Content}_{\text{diversity}}}$$

*Figure 6: Traditional Information Operation Impact Equation*

This equation encapsulates the core dynamics of time-honoured campaigns, accentuating the significance of operational outreach, intrinsic message quality, and the engagement metrics. The denominator embodies external challenges such as pervasive media interferences, inherent audience resistance, and other extraneous factors.

As we integrate AI capabilities, the model evolves:

$$\text{Impact}_{\text{AI}} = \frac{\text{Reach} \times \text{Adaptive Quality}(AI_{\text{factors}}) \times \text{Engagement}_{\text{AI}}}{(\text{Media Noise} + \text{Resistance} + AI_{\text{obstacles}}) \times \text{Content}_{\text{diversity}} - AI_{\text{advantages}}}$$

*Figure 7: AI-Augmented Information Operation Impact Equation*

Within this construct, real-time feedback refines the adaptive quality of the message, bolstered further by AI-driven engagement metrics. The AI spectrum introduces distinct strategic advantages, such as micro-targeting, which can dramatically curtail resistance and counter media disruptions.

Building upon this, with the specific linguistic prowess of LLMs, the equation refines further:

$$\text{Impact}_{\text{LLM}} = \frac{\text{Reach} \times \text{Adaptive Quality}(LLM_{\text{factors}}) \times \text{Engagement}_{\text{LLM}}}{(\text{Media Noise} + \text{Resistance} + LLM_{\text{constraints}}) \times \text{Content}_{\text{diversity}} - LLM_{\text{advantages}}}$$

*Figure 8: LLM and AI-Augmented Information Operation Impact Equation*

LLMs, with their unmatched capability in human-like narrative crafting and fine-tuned individual engagement, remarkably amplify the potential efficacy of information operations.

Yet, the intricate dance of strategic endeavours is not solely about efficiency. The moral compass and ethical bearings hold paramount significance. Thus, when guiding LLMs with ethical considerations, the equation translates to:

$$\text{Impact}_{\text{Ethical-LLM}} = \frac{\text{Reach} \times \text{Adaptive Quality}(LLM_{\text{ethical factors}}) \times \text{Engagement}_{\text{Ethical}}}{(\text{Media Noise} + \text{Resistance} + LLM_{\text{ethical constraints}}) \times \text{Content}_{\text{diversity}} - LLM_{\text{ethical advantages}}}$$

*Figure 9: Ethical LLM and AI-Augmented Information Operation Impact Equation*

This equation underscores the importance of ethically guided operations that align with societal norms and values, ensuring both the responsible and credible use of technology.

The "Nation-State Comparative Measure" equation offers a holistic perspective on the relative effectiveness of AI-augmented information operations against traditional strategies.

$$\text{Comparative Measure}_{\text{Nation-State}} = \frac{\text{Impact}_{\text{AI}} + \text{Impact}_{\text{LLM}} + \text{Impact}_{\text{Ethical-LLM}}}{\text{Impact}_{\text{Traditional}}}$$

*Figure 10: Nation-State Comparative Measure*

This measure, by encompassing the diverse facets of information campaigns from reach to real-time adaptability, provides a quantitative lens to assess the strategic shifts in the information warfare landscape.

As AI and LLMs become embedded in information operations, this comparative measure serves as a barometer for nation-states to gauge the net advantage (or otherwise) they achieve by transitioning from conventional methods. A value exceeding one indicates a favourable tilt towards AI-enhanced methods, whereas a value below one signals a need for introspection and strategy recalibration. This becomes particularly salient when considering the potential for AI and LLMs to inadvertently construct digital echo chambers, potentially intensifying societal schisms.

While the prior equations delve into the mechanics of information operations, the "Nation-State Comparative Measure" serves a strategic purpose, emphasising the necessity to consistently evaluate and iterate strategies in this dynamic domain. The inclusion of ethical considerations in these equations further cements the argument for a balanced approach, merging technological advancements with moral and ethical imperatives (Thuraisingham, 2020).

**Ethical and Strategic Considerations: AI Mediators in the Age of LLMs**

The ascendancy of Large Language Models (LLMs) in information operations

could pose challenges reminiscent of the nuclear dilemmas that defined the Cold War era (Mitra, 2023). These LLMs, while technologically astounding, navigate the vast information ecosystem devoid of the inherent moral and ethical compass so intrinsic to human judgement. The unparalleled power of nuclear capabilities in the past century (Mitra, 2023) has found its counterpart in the present: LLMs have the potential to influence, mislead, and mold perceptions on an unprecedented scale. The democratic bastions of the West, replete with their nuanced histories and imperfections (Lebow, 2007), are now tasked with forging ethical frontiers in this rapidly emerging domain.

LLMs, with their unmatched capabilities, operate in a vacuum of moral discernment. This absence of intrinsic ethics, amplified by their vast reach and speed, underscores the pressing need for robust ethical frameworks. Such frameworks should serve as the digital equivalents of "non-proliferation treaties", ensuring that LLM outputs resonate deeply with the democratic values, freedoms, and norms we cherish (Thuraisingham, 2020).

In an era where trust is anchored in transparency, it becomes essential to illuminate the arcane workings of these models. Stakeholders, ranging from the public to policymakers, merit insights into the training data, algorithms, and processes that drive these models, reinforcing accountability in this new age of digital persuasion (Welch & Fox, 2012).

The allure of LLMs is undeniable: unparalleled precision in messaging, adaptability in real-time, and scalability that dwarfs human capabilities. Yet, with profound power comes an even profounder responsibility. Historically, the democratic West has balanced power with accountability, leading to the pressing question: Just because we can, does it mean we should? The capabilities of LLMs might offer tantalising prospects of influencing narratives, but the ethical quandary persists: at what cost?

To harness the potential of LLMs while safeguarding the democratic tenets of the West, we need a three-pronged approach: robust ethical frameworks, unwavering human oversight, and responsive feedback mechanisms (Scharre, 2023). This journey mandates sustained dialogue among technologists, policymakers, ethicists, and society, ensuring that as we push the boundaries of technology, our moral compass remains steadfast.

Delineating the landscape are three pivotal equations:

**Unmitigated LLM Information Operations**:

$$\text{Impact}_{\text{Unmitigated-LLM}} = \frac{\text{Reach} \times \text{Content}_{\text{LLM}}}{\text{MediaNoise} + \text{Resistance} + \delta_{\text{CascadeEffects}}}$$

*Figure 11: LLM-augmented Information Operation with Cascade Effects*

Here, $\delta$CascadeEffects symbolises the unpredictable and potentially vast repercussions of uncontrolled LLM outputs. This equation underscores the potential pitfalls when LLMs operate devoid of checks, where unforeseen ripple effects could undermine the intended impact.

**Manual Ethical and Strategic Input**:

$$\text{Impact}_{\text{ManualInput-LLM}} = \frac{\text{Reach} \times \text{Content}_{\text{LLM}} \times \text{EthicalFactor}_{\text{human}}}{\text{MediaNoise} + \text{Resistance} + \gamma_{\text{Inefficiency}}}$$

*Figure 12: LLM-augmented Information Operation with Human Intervention*

In this paradigm, $\gamma$Inefficiency illustrates the inherent limitations of human intervention in overseeing LLM outputs. While human ethics might introduce a semblance of control, it does not fully exploit the potential of LLMs due to the intrinsic inefficiencies of manual oversight.

**AI Mediators in Action**:

$$\text{Impact}_{\text{AI-Mediator-LLM}} = \frac{\text{Reach} \times \text{Content}_{\text{LLM}} \times \text{EthicalFactor}_{\text{AI}}}{\text{MediaNoise} + \text{Resistance}}$$

This equation heralds the introduction of AI mediators (Wang, Xie, et al., 2023), acting as real-time ethical arbiters (Wang, Ma, et al., 2023), ensuring LLM outputs align with pre-defined ethical standards. The omission of inefficiency terms emphasises the merits of AI-driven oversight: the confluence of machine scalability with the ethical nuances of human values.

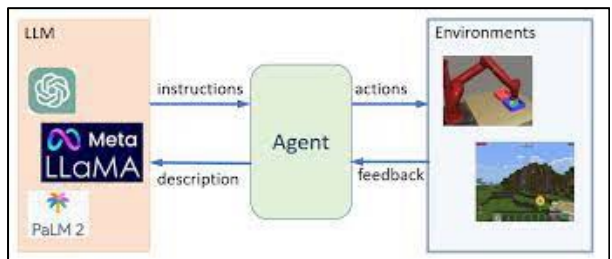

*LLMs and Agent for solving complex embodied tasks, Hu et al., 2023.*

Ethical pillars, oversight, and transparency aren't merely passive guidelines but should be intricately woven into the very essence of LLM operations (Zhou et al., 2023). By synergising LLMs with AI mediators (Wang, Xie, et al., 2023), mankind can edge closer to a future where technological advancements amplify, rather than erode, our values.

**Integrating Computational Social Science, Computational Ethics, Systems Engineering, and AI Ethics in LLM-driven Operations**

Society is charting unexplored territories, not just technologically but also ethically. As nation-state's harness these sophisticated models to potentially reshape information operations, it becomes imperative to approach this integration from a multidisciplinary lens. Four pillars serve as a guiding light: Computational Social Science, Computational Ethics, Systems Engineering Theory, and AI Ethics & Responsible AI Engineering.

At the heart of every information operation lies the individual—receiving, processing, and acting upon the information. Computational Social Science offers a lens to study these interactions at scale. As LLMs craft narratives, consideration of the broader societal networks in which these narratives propagate must be had. What are the cascading effects of a piece of information? How do group dynamics amplify or attenuate a message? By modelling these interactions, nation-states can predict and understand the ripple effects of LLM outputs, ensuring they align with broader societal objectives and do not unintentionally exacerbate divisions or biases.

Every piece of content generated by an LLM carries ethical weight. Computational Ethics provides the framework to embed these considerations directly into the algorithms. By integrating computational ethics, LLMs can operate within defined moral parameters, reflecting societal norms and values, while avoiding potential pitfalls such as misinformation or bias.

Systems Engineering Theory offers a comprehensive approach to LLM-driven operations. It emphasises the interconnectedness of all components—from the underlying algorithms to the AI agents monitoring outputs, to the societal systems in which the information operates. By understanding and modelling these interconnections, there can be optimisation of the entire system for both efficiency and ethical considerations, ensuring that the outputs align with strategic objectives, while minimising unintended consequences.

AI Ethics and Responsible AI Engineering bring forth the principle of accountability. As LLMs are deployed in high-stakes domains, there needs to be a clear framework for oversight, transparency, and redress. LLMs, despite their prowess, lack the intrinsic human moral compass (Vaswani et al., 2017). It is upon people to integrate this compass, ensuring that LLMs operate in a manner that's not just effective but also ethically sound, respecting individual rights and societal values.

By intertwining Computational Social Science, Computational Ethics, Systems Engineering, and AI Ethics, nation-states ensure that LLM-driven operations are not just technologically advanced, but also ethically grounded, societally beneficial, and holistically optimised.

**Looking Forward: "ClausewitzGPT"**

The integration of Large Language Models (LLMs) with AI agent mediators presents a groundbreaking approach to the arena of nation-state conflict information operations. This innovative methodology is proposed in the equation, henceforth known as "ClausewitzGPT":

$$ClausewitzGPT(P, S, T) = h\left(f_{LLM}(P), \sum_{i=1}^{n} f_{agent_i}(f_{LLM}(P), S, T), S, T\right) \times \omega_{machine-speed}$$

- *ClausewitzGPT*: This function captures the ethical and strategic effectiveness of the information operation.
- *P*: Represents the input prompt or directive to the LLM, symbolising the strategic message the nation-state aims to broadcast.
- *S*: Signifies real-world feedback, ranging from public sentiment metrics to counter-narratives, reflecting the dynamic backdrop of the operation.
- *T*: The time variable, tracing the progression and impact of the operation.
- $f_{LLM}(P)$: The mechanism by which the LLM generates content based on the input prompt.
- $f_{agent_i}$: The evaluation function of the $i^{th}$ agent, tasked with assessing the LLM's output for alignment with both strategic and ethical guidelines.
- $\omega_{machine-speed}$: The parameter that represents the speed, efficiency and exponential increase of a LLM-drive, AI-augmented information operation.
- *h*: A comprehensive function melding the LLM's content, AI agent decisions, real-world dynamics, and temporal factors to evaluate the operation's overall effectiveness.

In the hypothetical context of escalating geopolitical tensions, LLMs, guided by $f_{LLM}(P)$, could be strategically positioned, echoing Clausewitzian principles, to craft real-time propaganda and counter-narratives. However, resonating with Clausewitz's emphasis on moral forces and uncertainties in warfare, it's envisioned that AI agent mediators, utilising $f_{agent_i}$, would scrutinise these narratives for both their strategic intent and ethical alignment.

Diplomacy requires a careful hand. Theoretically, LLMs might generate nuanced statements, which AI agent mediators would then refine, aligning them with both strategic imperatives of the state and the ethical values of The Enlightenment.

Venturing into the speculative domain of intelligence and counter-intelligence, reminiscent of Clausewitz's 'fog of war', LLMs could decode and produce intelligence. In this imagined scenario, the AI agent mediators would act as the discerning 'eyes' (Wang, Xie, et al., 2023), ensuring clarity, strategic alignment, and an absence of misinformation, in the spirit of Clausewitz's emphasis on informed strategy.

Envisioning the role of LLMs in psychological operations, reflecting Clausewitz's contemplations on the moral dimensions of war (Clausewitz, 1832), these models might produce content aimed at influencing adversaries. The overseeing AI agent mediators, echoing Clausewitz's cautionary tones, would ensure that the content achieves its intended psychological effects without leading to unintended escalations (Williams et al., 2016).

Navigating the theoretical challenges posed by LLMs, one is reminded of Clausewitz's musings on the complexities of war. In this projected future, setting ethical boundaries in the age of AI would become paramount. With the potential to influence global narratives, my equation, inspired by the "ClausewitzGPT" framework, emphasises the critical balance between strategic objectives and ethical imperatives. As society contemplates potential AI-centric arms races in information warfare, the essence of Clausewitz's reflections on balancing power and principle becomes increasingly pertinent.

While the integration of LLMs and AI agent mediators remains within the realm of theoretical discourse (Zhou et al., 2023), their potential application could, in time, redefine information warfare, much as Clausewitz's doctrines reshaped military thinking. This synergy, encapsulated in the "ClausewitzGPT" equations, suggest a potential future where strategic narratives are both potent and ethically anchored.

## Conclusion

In the annals of information operations, the trajectory has been marked by consistent evolution, from rudimentary propaganda to sophisticated digital campaigns. Today, with the rise of LLMs, society stands at a pivotal moment, reminiscent of the great leaps in past communication revolutions. These LLMs, emblematic of the synthesis of artificial intelligence with strategic communication, have the potential to redefine information warfare.

Society has witnessed the surge of AI-augmented operations, where machines aid, amplify, and sometimes even autonomously drive narratives. "ClausewitzGPT" stands as a testament to this evolution, marrying classical strategic tenets with the prowess of LLMs. It represents a bridge between the wisdom of yesteryears and the technological marvels of today, guiding humankind towards a future where AI and strategy coalesce seamlessly.

The introduction of these novel mathematical equations and frameworks add precision to this domain; they also serve as navigational tools, ensuring that LLMs remains anchored to ethical and strategic considerations. These equations are a stark reminder: in the hands of the unprincipled, the power of LLMs can lead to an AI arms race, where the very fabric of reality becomes malleable.

This beckons the democratic institutions of the West, historically the torchbearers of Enlightenment values, to rise to the occasion. As the potential for an AI arms race looms, championing transparency, accountability, and ethical considerations is not just a responsibility—it's a moral imperative. The West has a unique opportunity to lay the foundations for responsible LLM use, ensuring as technology advances, the West remain rooted in the principles that have been the bedrock of their civilisations.

With LLMs, it's crucial to remember that tools, irrespective of their power, are only as good, as ethical, and as visionary as those who wield them.